\documentclass[conference,twocolumn]{IEEEtran}

\usepackage{amsmath,amsfonts,bm}

\def\eqref#1{equation~\ref{#1}}

\def\1{\bm{1}}

\DeclareMathAlphabet{\mathsfit}{\encodingdefault}{\sfdefault}{m}{sl}
\SetMathAlphabet{\mathsfit}{bold}{\encodingdefault}{\sfdefault}{bx}{n}

\usepackage{url}
\usepackage{hyperref}
\usepackage{graphicx} 
\usepackage{dirtytalk}

\usepackage[algo2e]{algorithm2e} 
\usepackage{algorithm}
\usepackage{algorithmic}

\usepackage{subfig}

\usepackage{array, multirow}
\usepackage{float}
\usepackage{tabularx}

\usepackage{amssymb}
\usepackage{makecell}

\ifCLASSINFOpdf
\else
\fi
\begin{document}

\title{Regime Identification for Improving Causal Analysis in Non-stationary Timeseries}
%
%
%

\author{
    \IEEEauthorblockN{Wasim Ahmad, Maha Shadaydeh, Joachim Denzler}
    \IEEEauthorblockA{ Computer Vision Group, Faculty of Mathematics and Computer Science, \\
    Friedrich Schiller University Jena, Germany
    \\wasim.ahmad, maha.shadaydeh, joachim.denzler}@uni-jena.de}


%



\maketitle

\begin{abstract}
Time series data from real-world systems often display non-stationary behavior, indicating varying statistical characteristics over time. This inherent variability poses significant challenges in deciphering the underlying structural relationships within the data, particularly in correlation and causality analyses, model stability, etc. Recognizing distinct segments or regimes within multivariate time series data, characterized by relatively stable behavior and consistent statistical properties over extended periods, becomes crucial. In this study, we apply the regime identification (\textbf{RegID}) technique to multivariate time series, fundamentally designed to unveil locally stationary segments within data. The distinguishing features between regimes are identified using covariance matrices in a Riemannian space. We aim to highlight how regime identification contributes to improving the discovery of causal structures from multivariate non-stationary time series data. Our experiments, encompassing both synthetic and real-world datasets, highlight the effectiveness of regime-wise time series causal analysis. We validate our approach by first demonstrating improved causal structure discovery using synthetic data where the ground truth causal relationships are known. Subsequently, we apply this methodology to climate-ecosystem dataset, showcasing its applicability in real-world scenarios.
\end{abstract}


%
\IEEEpeerreviewmaketitle

\section{Introduction}
\label{introduction} 

\begin{figure}[H]
\centering
\includegraphics[width=0.45\textwidth]{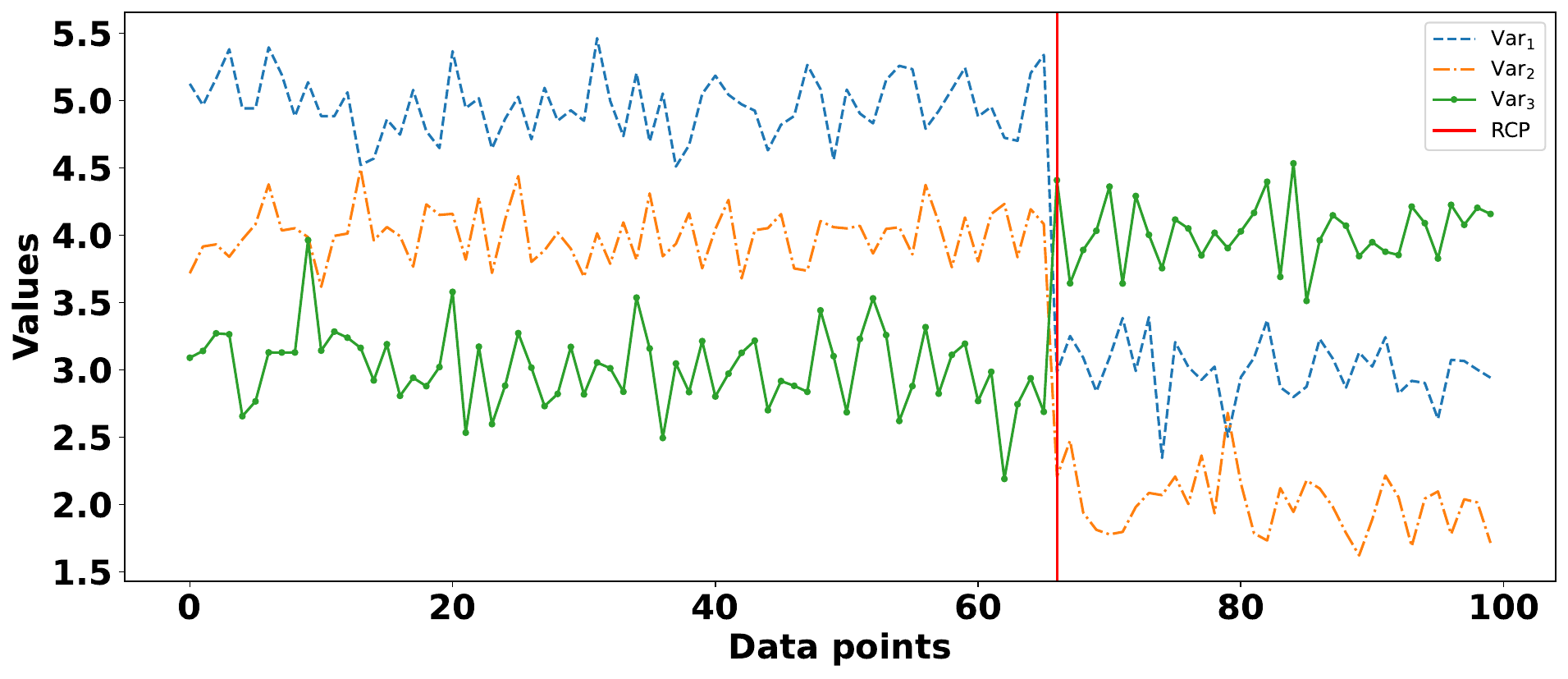}%
\caption{Illustration of regimes, i.e., distribution shift in time series over a period of time, separated by regime change point (RCP) in multivariate time series.} 
\label{fig:regdemo}
\end{figure}

Most often, the analysis and identification of the underlying system of multivariate time series is built on the stationarity assumption. However, the underlying dynamics may vary, e.g., over different seasons reflecting non-stationary behavior or changes in the system's state. Notably, there are shifts in patterns that give rise to distinct regimes, which typically refers to identifiable patterns with change points that the data exhibits over a significant period of time, demonstrated in Figure \ref{fig:regdemo}. The non-stationary behavior poses challenges in time series analysis, i.e., correlation and causality, model robustness, forecasting, etc. \cite{papana2023identification, arjovsky2019invariant}. A straightforward approach would be manually selecting segments instead of an entire time series for analysis. However, this disregards the stability and co-evolving dynamics inherent in multivariate time series. Hence, there is a need for automatic identification of regimes, which considers the stability and relationships among variables within the multivariate time series. In this work, we apply regime identification to obtain time series segments with stable statistical properties and enhance the understanding of the underlying causal structure in multivariate time series. In addition, identification of regimes or regime change points (RCP), i.e, points where regime switches, could be leveraged for various applications in climate (i.e., extreme events detection), health \cite{feuz2014automated, malladi2013online} and predictive maintenance \cite{zenisek2019machine}.

We achieve regime identification by clustering the covariance matrices of the multivariate non-stationary time series using \textit{k}-Means with Riemannian distance metric. This technique generates multivariate time series segments with locally consistent statistical properties. This idea has been previously applied in brain science to estimate functional connectivity in brain networks \cite{dai2019analyzing, slavakis2017riemannian, dodero2015kernel}, which we borrowed to climatic time series analysis. We performed experiments first on synthetic data as a proof of concept and show how coupling of RegID with the state-of-the-art causality methods CDMI \cite{ahmad2022causal}, VAR Granger causality (VAR-GC) \cite{granger1969investigating}, PCMCI$^+$ \cite{runge2020discovering} improves causal discovery. Subsequently we apply our approach to identify climate-ecosystem causal interactions by using RegID in tandem with , Vanilla-PC \cite{janzing2009telling}, Trace method \cite{zscheischler2012testing} and 2GVecCI \cite{wahl2023vector}.


\section{Related Work}
\label{section:related}

The work of \cite{tajeuna2023modeling} presents a method for regime shifts in co-evolving time series. They model multiple time series in an ecosystem by summarizing them into a mapping grid, capturing behavioral dependencies and regime transitions. The model utilizes a dynamic network representation for understanding time series behavior and employs a full-time-dependent Cox regression model for learning regime transitions. The work of \cite{truong2020selective} presents a comprehensive review of RCPs detection where they categorize the approaches into likelihood-based, kernel-based, and graph-based methods. 
The authors of \cite{ermshaus2023clasp} present ClasP, a parameter-free and domain-agnostic time series segmentation method that segments a uni-variate time series in an unsupervised manner. Methods like FLOSS \cite{gharghabi2017matrix} and ESPRESSO \cite{deldari2020espresso} use neural networks-based approaches for time series segmentation. Most methods do not consider the co-existence of the time series \cite{matsubara2019dynamic, chen2018neucast}. Indeed, examining multiple time series as an ecosystem implies a co-evolutionary dynamic among the series, aligning with the concept of complex systems and suggesting potential interactions between these variables at various time intervals. Investigating the interconnections among time series at different time intervals is essential for identifying and predicting regimes. The work of \cite{saggioro2020reconstructing} presents a regime-dependent causal structure from time series where they differ between two regimes based on the discovered causal structure with PCMCI \cite{runge2019detecting}. On the contrary, our approach is to estimate regimes based on covariance structures existing within the multivariate time series and discover the causal structure for each regime.

\section{Method}
\label{section:methods}
We use the time series's covariance matrix space to identify regimes previously used in brain science \cite{dai2019analyzing}, \cite{slavakis2017riemannian}, and \cite{dodero2015kernel}, where covariance matrices in Riemannian surfaces are used for estimating functional connectivity in brain regimes. The covariance structure can reveal how the multivariate time series change relative to each other, which makes it ideal for segmenting time series based on their dynamics and discovering the underlying causal structure. Let $Z(t)$ denote the multivariate time series at time $t$, and $w$ be the window size. We extract covariance matrices for each time window with no overlap until the end of the multivariate time series and obtain a pool of the covariance matrices over time $\{\Sigma(t)\}_{t=w}^{T}$ where $T$ is the total number of time steps in multivariate time series. We apply \textit{k}-means with Riemannian and Euclidean distance metrics to group non-stationary time series based on their dynamics.

The \textit{k}-Means clustering using Euclidean distance is represented as:
\begin{equation}
\label{eqn:euclid}
C_{\mathcal{E}} = \text{argmin}_{C} \sum_{t=w}^{T} \sum_{i \in C} \left\| \Sigma(t) - \mu_i \right\|^2    
\end{equation}
where \( C \) represents the set of cluster indices, \( \mu_i \) is the centroid of cluster \( i \), and \( \left\| \cdot \right\| \) denotes the Euclidean norm. Similarly, for the \textit{k}-Means clustering Riemannian distance, the objective function is represented as:
\begin{equation}
\label{eqn:reiman}
    C_\mathcal{R} = \text{argmin}_{C} \sum_{t=w}^{T} \sum_{i \in C} d(\Sigma(t), \mu_i)^2
\end{equation}
where \(d(\cdot, \cdot) \) is the Riemannian distance using the Log-Euclidean, an affine-invariant metric that remains unchanged under affine transformations: $d(\Sigma_i, \Sigma_j) = \left( \sum_{i=1}^{n} \log^2(\lambda_i) \right)^{1/2}$.
Here \( \lambda_i \) represents the eigenvalues of the matrix \( \Sigma_i^{-1/2} \Sigma_j \Sigma_i^{-1/2} \). These eigenvalues are used to compute the Riemannian distance between two symmetric positive definite matrices (SPDMs) \( \Sigma_i \) and \( \Sigma_j \). Riemannian is a distance measurement that deals with the study of nonlinear or curved surfaces. Unlike Euclidean, which measures the straight-line distance between two points on a flat surface, Riemannian distance represents the shortest path along a curved surface where the geometry of the space is considered. The pseudo-code for regime identification in multivariate non-stationary time series is given in Algorithm \ref{algo:regime}.

\begin{algorithm}
\begin{algorithmic}
\STATE \textbf{function} $\textsc{regimes}(Z, w, k, dim)$
\STATE \textbf{Input:} $Z_{i, i = 1, \dots, N}$ is N-variate non-stationary time series, $w$ is the window size. If k (number of expected regimes) is not provided, then optimally determined by algorithm.
\STATE \textbf{Output:} $\mathcal{R}$ are identified regimes in non-stationary time series. 

\STATE \textbf{for} $j \gets 1$ to $l/w$ \newline
\tcp{$l$ is the length of the time series}
\STATE \hspace{30pt} $\Sigma_j \gets \text{covariance}(Z, w)$
\STATE \hspace{30pt} spdms[j] $\gets \Sigma_j$
\STATE \textbf{end for}
\STATE \textbf{if} dim != full: 
    \STATE \hspace{30pt} spdms $\gets$ reduce$_{dim}(spdms, n)$
\STATE \textbf{if} k is \textbf{None}: 
    \STATE \hspace{30pt} k = optimal$_k$(Z) 
    
    \tcp{based on the elbow point of the Calinski-Harabasz score curve}
\STATE $R \gets \textit{k}\text{-Means}(spdms, k)$ \newline
\tcp{Clustering  of covariance matrices or spdms using Riemannian/Euclidean as distance metrics}
\STATE \textbf{Return} $\mathcal{R}:\{R_{r, r=1, ..., k}\}$
\end{algorithmic}
\caption{: \textbf{RegID}}
\label{algo:regime}
\end{algorithm}

\begin{figure*}
    \centering
    \subfloat[\centering Euclidean]{{\includegraphics[width=0.45\textwidth]{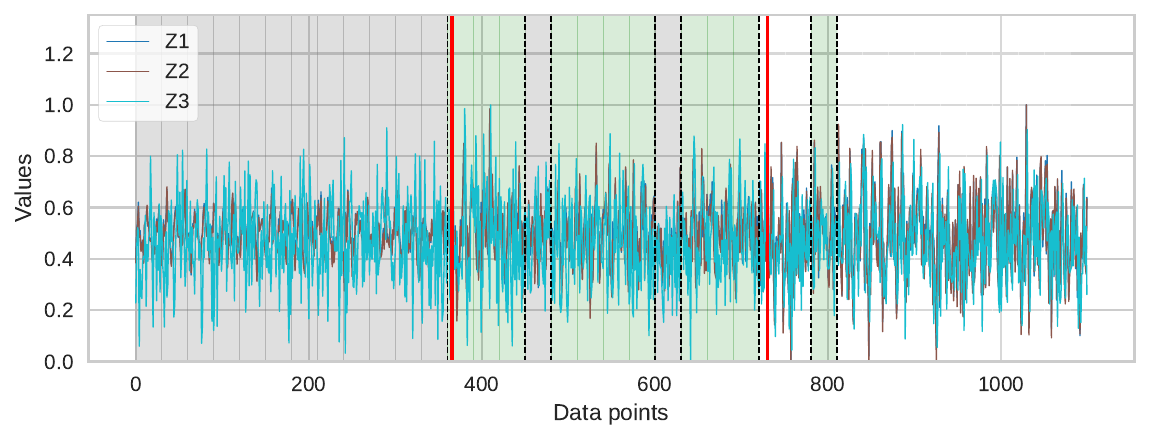}}}
    \subfloat[\centering Riemannian]{{\includegraphics[width=0.45\textwidth]{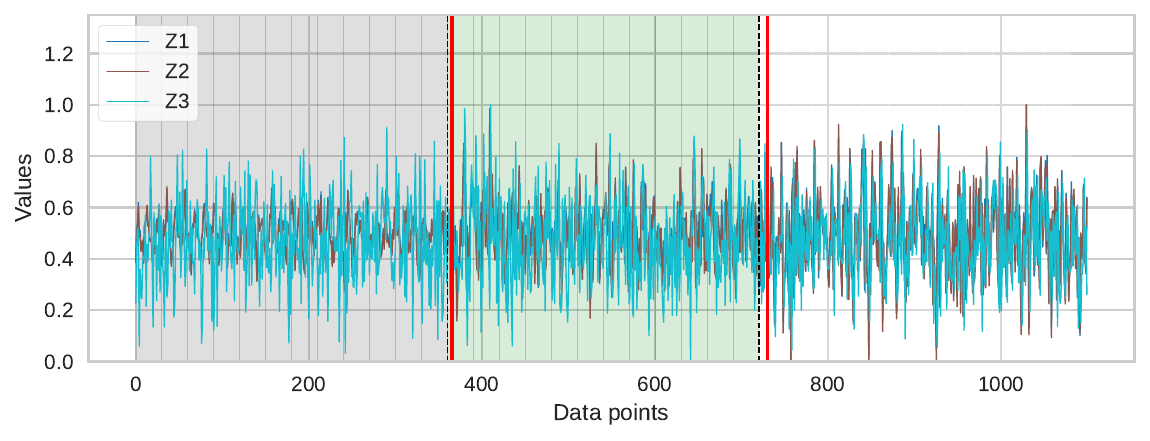}}}
    \caption{Detected regimes using \textbf{a}. Euclidean and \textbf{b}. Riemannian distance metric, are separated by black dotted lines while ground truth RCPs are shown in red lines.}
    \label{fig:regime_paper}
    \vspace{-1mm}%
\end{figure*}

\section{Experiments}
\label{section:experiments}
\paragraph{Synthetic Data} To evaluate the performance of our method, we use synthetic data model: $\label{eqn:syndata}
Z_{j, t} = a_{j}Z_{j, t-1} + \Sigma_i c_i f_i (Z_{i, t-\tau_i}) + \eta_{j, t}$. The system variables $Z_j$ have auto and cross-functional dependencies with a time delay of  $\tau$. The data model incorporates linear and nonlinear dependencies $f$, i.e., exponential, polynomial, and adds uncorrelated, normally distributed noise $\eta_t^j$. We artificially inject regimes in the generated multivariate time series. To evaluate the effectiveness of the regime identification method, we conducted experiments using artificially generated non-stationary time series. Changes in statistical properties such as mean, variance, and noise were introduced at specific time points to simulate different regimes. We embedded three distinct regimes within the data, each marked by specified RCPs. 

Figure \ref{fig:regime_paper} and \ref{fig:synregimes} illustrates the qualitative outcomes of regime identification in the synthetic time series employing \textit{k}-Means with both Riemannian and Euclidean distance measures. The regime identification process utilizing the Riemannian metric closely aligns with the ground truth by considering the nonlinear geometry of covariance matrix-associated surfaces, which the Euclidean metric method overlooked. Similar findings have been reported in the work of \cite{kainolda2021riemannian}. Riemannian has advantages over Euclidean as a distance metric in the analysis of covariance matrices. First, it considers the geometry of the space, which is essential when comparing data represented as SPDMs. Besides, it is invariant to data transformations, i.e., scaling, rotations, and translations. Moreover, it fulfills the triangle inequality property, which states that the distance between any two points is always less than or equal to the sum of the distances between those points and a third point. This property is vital for distance-based clustering and classification methods as it allows for efficient computation of distances and ensures that the resulting clusters are coherent and well-separated, as we have seen in our experiments. For regime identification in synthetic multivariate time series utilizing covariance structure in a Riemannian space demonstrates greater stability across various sizes of time series batches. In contrast, the Euclidean metric is highly sensitive to changes in window size and generally fails to accurately identify regimes in non-stationary time series, except for a window size of 90 in our experiments, see Figure \ref{fig:synregimes}. The red vertical lines in the figures indicate the true RCPs in the dynamics of the time series. We segmented the time series by varying window sizes from 15 to 90, with detected regimes shaded in different colors to distinguish one regime from another.

\begin{table}[H]
\centering
\small
\caption{Performance analysis of causality methods on synthetic time series with and without regime identification. For regime-wise causal discovery, the mean values over all regime are shown.}
\begin{tabular}{lllll}
    \hline
    Methods & Precision & Recall & Accuracy & F-score \\
    \hline
    \hline
    CDMI & 0.60 & 0.90 & 0.72 & 0.72 \\
    \cline{2-5}
    RegID-CDMI  & 0.96 & 0.87 & 0.93 & 0.91 \\
    \hline
     VARGC & 0.50 & 1.00 & 0.60 & 0.67 \\
      \cline{2-5}
     RegID-VARGC  & 0.72 & 0.90 & 0.77 & 0.79 \\
    \hline
     PCMCI$^+$ & 0.58 & 0.70 & 0.68 & 0.63 \\
    \cline{2-5}
      RegID-PCMCI$^+$ & 0.69 & 0.73 & 0.76 & 0.71 \\
     \hline
    
\end{tabular}
\label{tab:synregresults}
\vspace{-1mm}%
\end{table}

In Table \ref{tab:synregresults}, we provide a comparative analysis of the performance of the CDMI, VAR-GC, and PCMCI$^+$ methods with and without regime identification. The regimes exhibit locally stable statistical properties, where the causality methods generally demonstrate superior F-score and accuracy (values are averaged over all identified regimes) compared to their performance without regime identification. This outcome is anticipated since the CDMI method relies on modeling nonlinearity in a stationary system, where both the learning and testing data originate from similar distributions, a condition satisfied within each regime. VAR-GC and PCMCI$^+$ also exhibit enhanced regime-wise performance. All causality methods assuming data stationarity are expected to show improved performance in causal discovery analysis when preceded by regime identification.

\begin{figure}[H]
\centering
\includegraphics[width=0.5\textwidth]{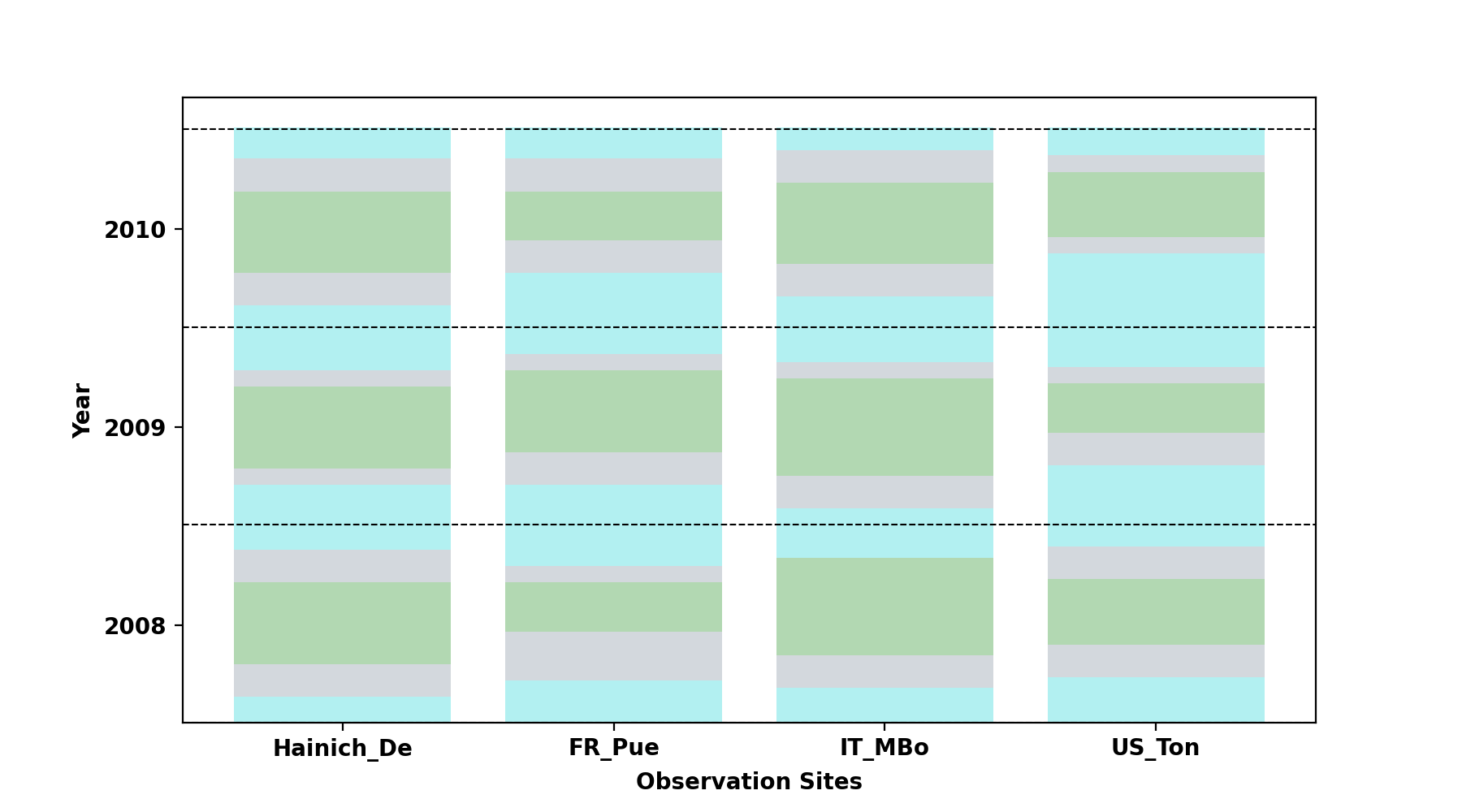}%
\caption{Identified regimes (blue, gray, green) in various sites in FLUXNET dataset where summer and winter periods are clearly separated by transition phase in each year.} 
\label{fig:sitecomp}
\end{figure}

\begin{table*}
\centering
\small
\caption{Performance of the methods in identifying causal direction (fractional occurrence) in climate-ecosystem data across regimes for various sites. We test a bi-directional causal link only for CDMI as it facilitates so.}
\begin{tabular}{llllll}
    \hline
     & & \multicolumn{4}{c}{Methods}\\
     \cline{3-6}
     Sites & Links & CDMI & 2GVecCI & Trace & Vanilla-PC \\
    \hline
    \hline
    \multirow{2}{*}{\makecell{DE-Hai \\ (Deciduous Broadleaf \\ Forests)}} & $G_{\text{C}} \rightarrow G_{\text{E}}$ & 0.17 & 0.66 & 0.50 & 0.00 \\
    \cline{2-6}
    & $G_{\text{C}} \leftarrow G_{\text{E}}$  & 0.00 & 0.17 & 0.50 & 0.17 \\
    \cline{2-6}
    & $G_{\text{C}} \leftrightarrow G_{\text{E}}$  & 0.83 & $-$ & $-$ & $-$  \\
      \cline{2-6}
    & $G_{\text{C}} \nleftrightarrow G_{\text{E}}$  & 0.00 & 0.17 & 0.00 & 0.83 \\
    \hline
    \multirow{2}{*}{\makecell{IT-MBo \\ (Grasslands)}} & $G_{\text{C}} \rightarrow G_{\text{E}}$ & 0.66 & 0.34 & 0.83 &  0.00 \\
    \cline{2-6}
    & $G_{\text{C}} \leftarrow G_{\text{E}}$  & 0.17 & 0.00 & 0.17  & 0.00 \\
    \cline{2-6}
    & $G_{\text{C}} \leftrightarrow G_{\text{E}}$  & 0.17 & $-$ & $-$ & $-$  \\
    \cline{2-6}
    & $G_{\text{C}} \nleftrightarrow G_{\text{E}}$  & 0.00 & 0.66 & 0.00 & 1.00 \\
    \hline
    \multirow{2}{*}{\makecell{FR-Pue \\ (Evergreen Broadleaf \\ Forests)}} & $G_{\text{C}} \rightarrow G_{\text{E}}$ & 0.34 & 0.34 & 1.00 & 0.50 \\
    \cline{2-6}
    & $G_{\text{C}} \leftarrow G_{\text{E}}$  & 0.00 & 0.00 & 0.00 & 0.17 \\
     \cline{2-6}
    & $G_{\text{C}} \leftrightarrow G_{\text{E}}$  & 0.66 & $-$ & $-$ & $-$  \\
    \cline{2-6}
    & $G_{\text{C}} \nleftrightarrow G_{\text{E}}$  & 0.00 & 0.66 & 0.00 & 0.33  \\
    \hline
     \multirow{2}{*}{\makecell{US-Ton \\ (Woody Savannas)}} & $G_{\text{C}} \rightarrow G_{\text{E}}$ & 0.34 & 0.34 & 1.00 & 0.50 \\
    \cline{2-6}
    & $G_{\text{C}} \leftarrow G_{\text{E}}$ & 0.00 & 0.00 & 0.00 & 0.17\\
     \cline{2-6}
    & $G_{\text{C}} \leftrightarrow G_{\text{E}}$  & 0.66 & $-$ & $-$ & $-$  \\
    \cline{2-6}
    & $G_{\text{C}} \nleftrightarrow G_{\text{E}}$ & 0.00 & 0.66 & 0.00 & 0.33 \\
    \hline
    
\end{tabular}
\label{tab:realdataresults}
\vspace{-1mm}%
\end{table*}

\paragraph{Climate Data}

Here, we conduct a causal analysis of environmental time series data following the identification of different regimes. We perform experiments using the FLUXNET2015 dataset\cite{pastorello2020fluxnet2015}. This dataset, acquired through the eddy covariance technique, captures carbon, water, and energy cycling between the biosphere and atmosphere across multiple regional networks. Data preparation efforts are carried out at individual site and network levels. For our experiments, we selected focus on several measurement sites, including Hainich (DE-Hai), Monte Bondone (IT-MBo: Grasslands), Puechabon (FR-Pue: Evergreen Broadleaf Forests), and Tonzi Ranch (US-Ton: Woody Savannas). We selected these sites due to their diverse ecological characteristics, representing different ecosystem types across different geographical regions. The FLUXNET2015 dataset encompasses climatic and ecological time series data, such as global radiation (\(R_\text{g}\)), temperature (\(T\)), gross primary production (\(GPP\)), and ecosystem respiration (\(R_{\text{eco}}\)), measured at various time scales (e.g., half-hourly, hourly, daily, weekly). We categorize these variables into two groups: the climate group (\(G_{\text{C}}\)), comprising temperature (\(T\)) and global radiation (\(R_\text{g}\)), and the ecosystem group (\(G_{\text{E}}\)), consisting of the ecosystem variables \(GPP\) and \(R_{\text{eco}}\). We opt for daily sampling frequency, which offers advantages in mitigating the influence of daily patterns that might obscure the underlying causal relationships in the data. Our results, presented in Table \ref{tab:realdataresults}, presents the fractional occurrences of the specified causal links from climate-ecosystem interactions across multiple sites and various regimes from 2008 to 2010 as shown in Figure \ref{fig:sitecomp}. The CDMI method shows a notable prevalence of bidirectional causal links ($G_{\text{C}} \leftrightarrow G_{\text{E}}$), suggesting robust mutual interactions between climate and ecosystem variables specifically in forest sites. In contrast, the 2GVecCI method effectively identifies causal links ($G_{\text{C}} \rightarrow G_{\text{E}}$) primarily in the DE-Hai site, but struggles to detect such links in other sites. The Trace method consistently highlights causal links ($G_{\text{C}} \rightarrow G_{\text{E}}$) across all sites, along with CDMI for grasslands. Conversely, Vanilla-PC demonstrates limited success in identifying climate-ecosystem links, with notable detection only in the US-Ton site (fractional occurrence of 0.5).



\section{Conclusion}
\label{section:conclusion}

In conclusion, our study introduces the coupling of regime identification (RegID) technique with causality methods as a powerful approach for improving causal analysis by revealing locally stationary segments within multivariate time series data. The challenges in time series analysis, i.e, correlation and causation, posed by non-stationary behavior in real-world systems, including changing statistical characteristics over time, are addressed through the identification of distinct regimes characterized by stable behavior and consistent statistical properties. Leveraging the covariance properties of the data and employing principles from Riemannian geometry, RegID effectively extracts distinguishing features among these regimes. Our experiments demonstrate the efficacy of the regime identification technique in enhancing the identification of distinct regimes and improving causal analysis of the non-stationary time series. As a future work, we conduct experiments on regime identification for high-dimensional data in a reduced-dimension space.



\section*{Acknowledgments}

This work is funded by the German Research Foundation (DFG) research grant SH 1682/1-1 and the Carl Zeiss Foundation within the scope of the program line \protect\say{Breakthroughs: Exploring Intelligent Systems} for \protect\say{Digitization — explore the basics, use applications}.

\ifCLASSOPTIONcaptionsoff
  \newpage
\fi



%


\appendix
\renewcommand\thefigure{\thesection.\arabic{figure}}  
\onecolumn
\setcounter{figure}{0} 

\begin{figure}[H]
\centering
\includegraphics[width=0.88\textwidth]{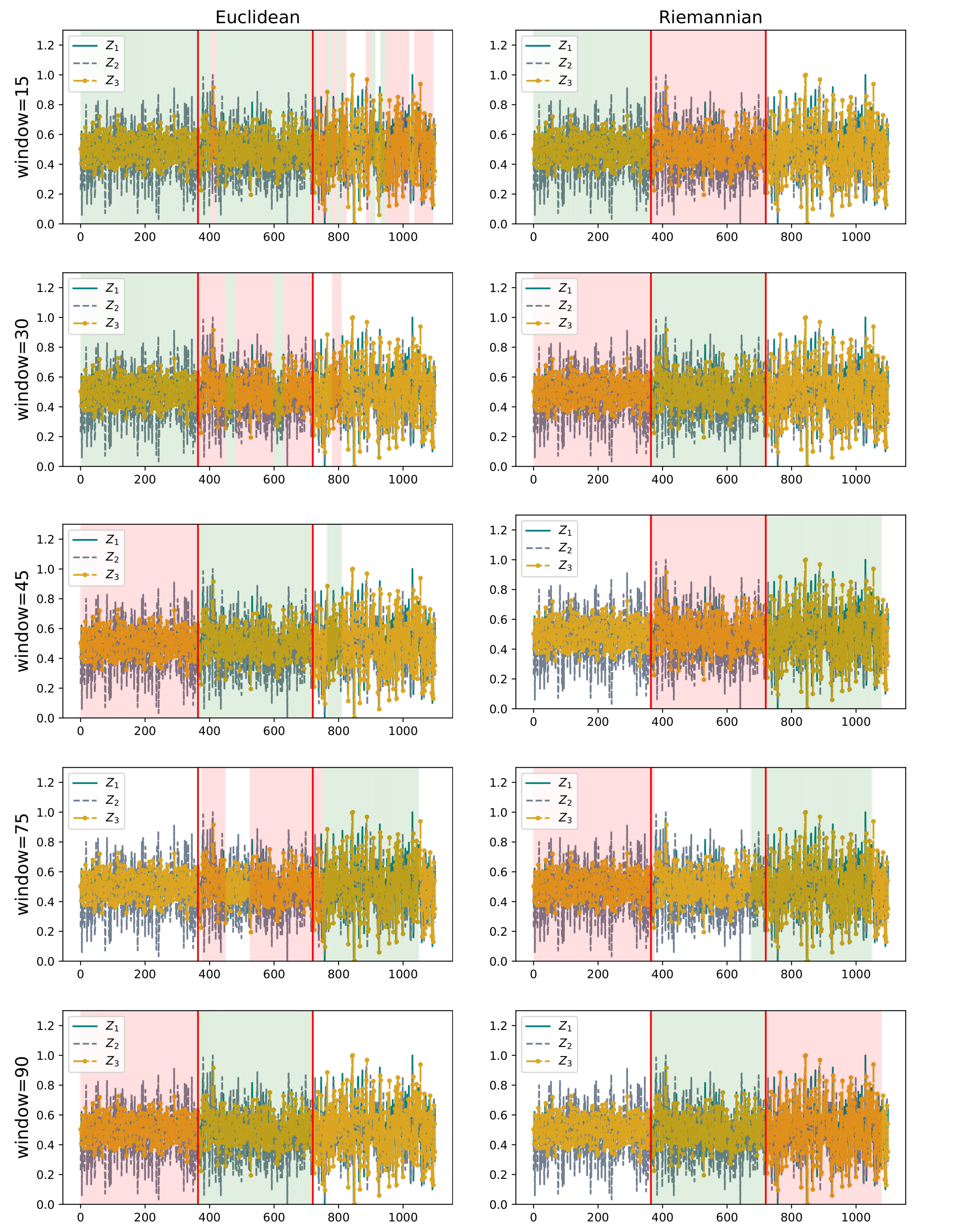}
\caption{Displayed are the identified regimes in synthetic time series $Z$ using \textit{k}-means with Euclidean and Riemannian metrics for a variety of window sizes [15-90]. The identified regimes are shown in red, green, and white. The red vertical lines show the actual change points.}
\label{fig:synregimes}
\end{figure}
\end{document}